MULTISCALE MODELING OF PERITECTIC REACTION GROWTH

Marcel Ausloos

SUPRATECS[*], B5, University of Liège, B-4000 Liège, Euroland

Rudi Cloots

LCIS[†], Institute of Chemistry, B6, University of Liège, B-4000 Liège, Euroland

Nicolas Vandewalle

GRASP[‡], Institut de Physique B5, Université de Liège, B-4000 Liège, Belgium

Peritectic growth of bulk superconducting cuprate perovskite, like $YBa_2Cu3O_{7-x}$ has been studied. The microstructure features are recalled. The simplest model giving the best description of the microstructure complexity at the mesoscopic scale is an Eden model adapted to chemical reactions. A repulsive dynamical interaction between the front and the solid 211 particles, rejected by the solidifying matrix, is also taken into account. A growth probability-with chemical reaction and pushing -transfer matrix method is developed. The simulations put into evidence the effect of the initial 211 size distribution on the microstructure.

1.   INTRODUCTION

---

[*]SUPRATECS = Services Universitaires Pour la Recherche et les Applications Technologiques de matériaux Electro-céramiques, Composites, Supraconducteurs

[†]LCIS = Laboratoire de Chimie Inorganique Structurale, a member of the SUPRATECS Centre

[‡] GRASP = Group for Research in Applied Statistical Physics

A huge step in material science was the discovery of high $T_C$ superconducting ceramics of the YBaCuO type, the so called 123 phase by Chu et al.[1]. Much improvement in the crystal growth process understanding is still needed. Indeed these ceramics have a complex microstructure, are brittle, susceptible to corrosion, present cracks, and are not easily prepared even at the laboratory stage. They are obtained through a peritectic reaction.

A peritectic reaction is an isothermal, reversible reaction between two phases, a liquid and a solid, that results, upon cooling of a binary (ternary, ... , *n* system) in one, (two, ... *n*-1) new solid phases. The reaction process is synonymous with "incongruent reaction". In the most simple case, a solid phase ($\alpha$) and a liquid (*L*) phase will together form a second solid phase ($\beta$) at a particular temperature and composition, such that $L + \alpha \rightleftharpoons \beta$.

These reactions are rather sluggish as the product phase will form at the boundary between the two reacting phases thus separating them, whence slowing down any further reaction. Peritectics are not as common as eutectics and eutectiods, but do occur in many alloy systems. Several cases of binary compounds having peritectic points in their phase diagrams are well known, like CaAl, YAl, BC, FeC, WC, ….

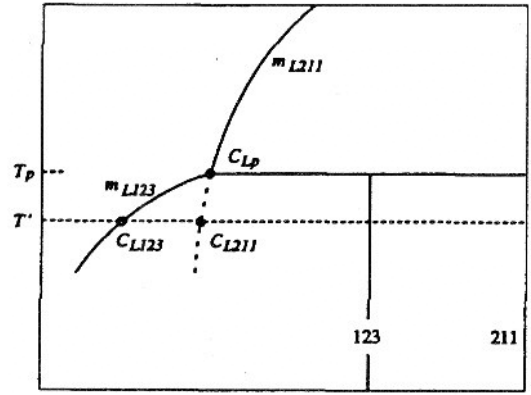

Fig. 1 Sketch of phase diagram near the peritectic point of the YBCO 123-211 competitive grow phases

In the following we concentrate on the case of 123-YBCO production, in order to make things more precise. Fig. 1 is a sketch of the phase diagram region where the peritectic reaction of interest occurs. This diagram is a cut through the ternary phase diagram $BaO/CuO_x/(1/2)Y_2O_3$ along the line joining the 123 and 211 phases, i.e. $YBa_2Cu_3O_7$ and $Y_2BaCuO_5$ or here above $\beta$ and $\alpha$, in the chemical reaction, respectively. The control of the density, the size distribution and the spatial distribution of the 211 particles are very relevant features characterizing bulk superconducting 123 materials and their processing.

a)

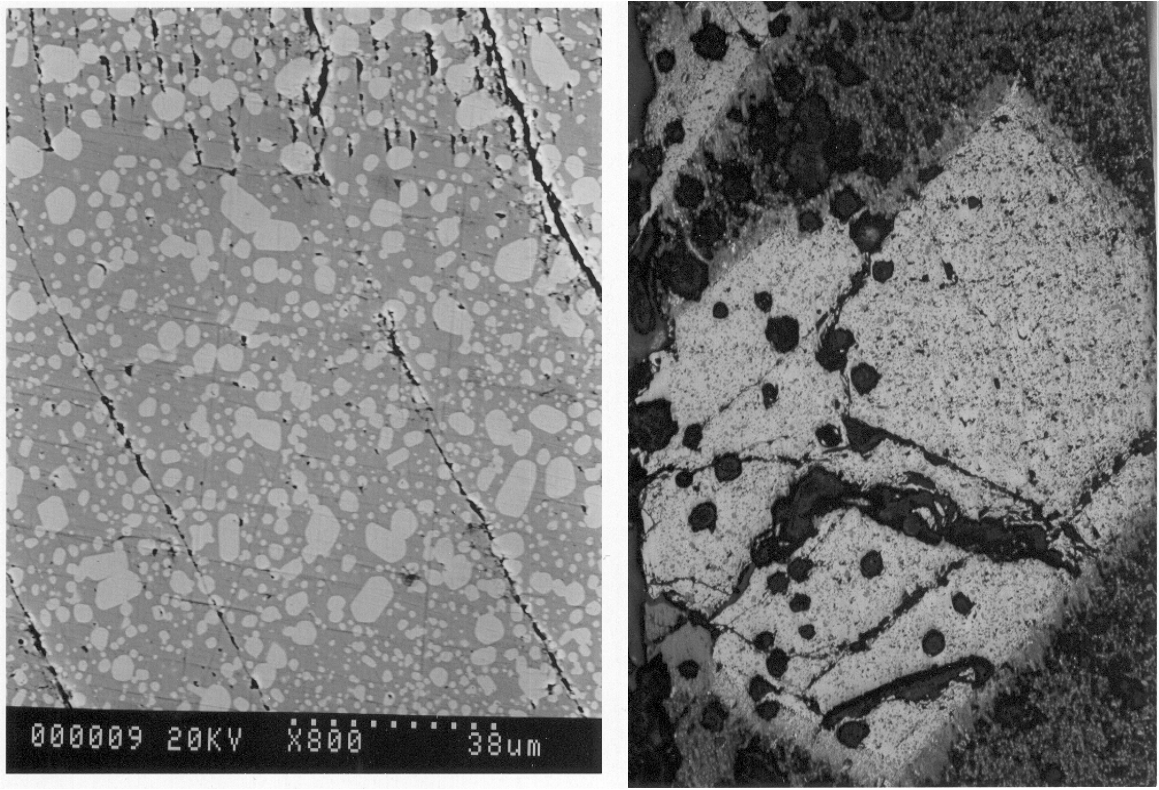

b)

Fig. 2 (a) SEM of 211 particles in a 123 matrix; (b) polarized light picture of a region with a grain boundary, indicating a different distribution of 211 particles in the 123 grain and outside (in the quenched liquid)

2. EXPERIMENTAL

Some powder mixture of starting oxides is quickly raised from room temperature above the melting line $m_{L123}/m_{211}$ (Fig. 1) which has a break at the peritectic temperature $T_p$ (about 1010°C) and a concentration $c_{Lp}$. Upon slow cooling down to $T_p$ the system decomposes into a liquid phase and a 211 solid phase. Below $T_p$, a 123 phase appears around nuclei and grows leading to some mixture of phases rich in 123 but still containing 211 particles. The yttrium ions needed for the growth of the 123 superconducting phase are provided by the dissolution of the 211 particles in the liquid.

It is of course impossible to correctly hit the 123 composition since at $T_p$ the system separates into the 211 phase and the $c_{Lp}$ mixture. Upon cooling, there are necessarily 211 particles trapped into the 123 phase and the residual matrix and/or located at the grain boundaries[2]. In fact, much work has been done on the study of the microstructural features presented by such melt-textured $YBa_2Cu_3O_{7-\delta}$ and related compounds[3,4], - the reference list being huge.

Nucleation is supposed to take place spontaneously, though heterogeneously, at the processing temperature, taken sligthly below ($\approx$ 5°C) $T_p$. One could remark that the number of nuclei can be experimentally tuned by modifying the cooling rate and the starting mixture composition. After the nucleation process is initiated, the temperature is slowly decreased at a rate about 1°C/h down to 950°C such that the process is *quasi* isothermal. In so doing, some 211 particles are trapped by the solidification front, but are also pushed along the interface. Thereafter, the polycrystalline structure is wholly solidified and a faster cooling rate can be applied down to room temperature. This can result in inhomogeneous 211 spatial distributions in the microstructure, as shown in Fig. 2 (a-b). In addition to the *L* and 123 phase contents, the spatial distribution of trapped particles is thus an important signature of the process[5,6,7]. It is usually found that at low initial 211 particle concentration, the particles are preferentially segregated at grain corners on terraces of bipyramids. In the case of a high initial concentration of 211 particles, these can be segregated in triangular or trapezoidal regions (depending on the orientation of the grain).

The SEM picture (Fig. 2 a) presents a cut through a planar growth front showing the 123 phase (in grey) structure. One can see many 211 particles (in light grey) trapped in the 123 matrix. Observe in Fig. 2 b that the density of 211 particles in the 123 phase is effectively lower than in the melt, outside the grain planar front, a difference which illustrates the dissolution of a large fraction of 211 particles. Such a high density of 211 particles is usually observed in the melt, more exactly at the grain border (Fig. 2 b).

3. MODEL

A modern approach to multiphase growth description can be found in phase field methods; see these proceedings, or in Schmitz et al.[8] specifically. It is, however, quite difficult to include together chemical reactions, particle motion and size effects in such algorithms. We have thus proposed another type of so called scale free (or multiscale) algorithm to simulate the microscopic features and later suggest practical considerations in growing large materials. The main features of the model, whence algorithm, combines a mechanism for pushing and/or trapping of particles, a chemical reaction, and crystal growth kinetics. It can be argued that the model is an adapted Eden model[9-12].

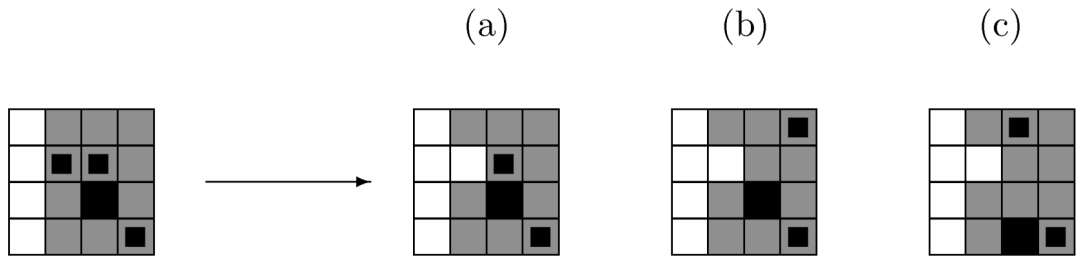

Fig. 3 Sketch of three (a-c) grow rules with the different types of site occupation

A theory of particle trapping or displacement along growing interfaces has been elaborated by Uhlmann et al, and is called the UCJ model[13]. The particles are either trapped or pushed by a growing solid/liquid interface depending on (i) the surface tensions at solid/liquid and particle/liquid interfaces, (ii) the particle size and (iii) the growth velocity of the solidifying front. For a fixed particle size, the particles are trapped by the front if the growth velocity of the interface is higher than a critical value. This critical value is controlled by the particle size and the interfacial tension energies. If small 211 particles are "less easily" trapped, they "more easily" move along the solid/liquid interface. However in some regions, the 211 particles are consumed by the peritectic recombination. On the other hand, the dissolution of 211 particles followed by a recombination process is not always complete.

The (multigrain) growth is considered as a discretized process in both space and time[14-19]. The model is mesoscopic in the sense that each mesoscopic cell of the lattice can receive a solid or a liquid unit. Different 211 particle size initial concentrations have been investigated. Each cell can contain (i) a liquid unit of phase *L*, (ii) a 123 crystal unit, (iii) a small 211 particle embedded in some melt or (iv) a 211 solid unit representing a large 211 particle (Fig. 3). Initially, each cell of the lattice contains either a large 211 particle with a probability $l_{211}$, or a small 211 particle embedded in some liquid *L* with a probability $s_{211}$, or a liquid unit with a probability $1 - l_{211} - s_{211}$.

Three different situations (Fig. 3) occur for the growth of a 123 unit:

(a) The dissolution of a small 211 particle in the L *on* the growing site leading to a new 123 unit,

(b) The dissolution of a small 211 particle *in the neighborhood* of the growing site (beside a 123 growth on the hit site),

(c) In addition to (a) and (b) the repulsion of a large 211 particle *toward a site in the neighborhood* of the growing site.

Nucleation is induced by simultaneously turning a number *n* of small 211 particles into initial 123 solid units, supposed to be randomly dispersed in the initial melt.

Since the process occurs over long time scales, thermodynamic (i.e. equilibrium-like) quantities can be mapped into multigrain growth probability rules, as follows. At each growth step, all *i* cells containing some liquid phase, i.e. *L* cells and cells with a small 211 particle, in contact with 123 mesoscopic cells are selected. The probability $P_i$ to grow the 123 phase on the cell *i* is given by a classical thermodynamic argument as

$$P_i \sim exp(-\Delta G_i/kT),$$

where $\Delta G_i$ is the relevant gain of free energy. Usually it can be decomposed into two terms: a bulk contribution depending on the driving force and a local surface contribution which is proportional to a chemical bond energy $J$. Since the bulk contribution is roughly constant in the system for isothermal conditions, only the local surface contribution is needed for measuring the probability of growth, i.e. $P_i \sim \exp^{-g_{nn}N_i}$ where $N_i$ is the number of nearest-neighboring (*nn*) 123 units belonging to the same grain and $g_{nn} = J_{nn}/(kT)$. This expresses an anisotropic locally preferred kinetic growth along the main directions. For high positive values of $g_{nn}$, square-like grains are growing[16,17]. When $g_{nn}$ is set to 1000, smooth grain faces are obtained. From a kinetic point of view, it is accepted that the 123 grains grow with a four-fold symmetry along the $a_{123}$ - $b_{123}$ planes, the edges of the grains being oriented along the (10) and (01) crystallographic directions. This has been implemented in the choice of the parameter value $g_{nn}$. This algorithm leads to a growth (G) transfer matrix-like description for lattice site occupation.

In order to simulate best the chemical reaction process, the mesoscopic cells for which the nearest and next-nearest neighbors do not contain any 211 particle are excluded from the growth cell selection. Indeed, the growth of the 123 phase on these mesoscopic cells is assumed to be improbable because of the local deficiency of yttrium around these cells. Taking into account the probabilities $P_i$ of all possible growing cells, one specific growth site is randomly selected. The peritectic reaction takes place in this cell through the (partial or total) dissolution of a 211 particle located on the chosen cell *or* in the neighborhood of the chosen cell. This algorithm leads to a chemical reaction (R) transfer matrix-like description for lattice site occupation.

Three different dynamical versions of the multigrain growth model have been considered from the particle trapping/pushing point of view. Only the case of nearest neighbor particle collisions have been so far considered. First (model I), all 211 particles can remain static in the melt. When two 211 particles are touched by the solid 123 front: (a) the position of both particles remains unchanged (as in model I); (b) the small particle jumps to a neighboring liquid site in order to reduce its number of neighboring 123 unit contacts (model II); (c) both particles jump to neighboring liquid sites in order to reduce their number of neighboring 123 unit contacts (model III).

In model (II), the small 211 particles are mobile when pushed by the growing grain while the large 211 particles are still static. This means that a small 211 particle touched by a 123 growing grain tends to reduce the front contact through a small jump onto a nearest or also to a next-nearest neighbor liquid site. If the particle cannot reduce its 'contact number' with neighboring 123 cells, the particle remains still for ever and is engulfed by the 123 phase for the next growth steps. This rule introduces a dynamical and repulsive interaction between the growth front and the small 211 particles. Notice that the small 211 particles are allowed to move sideways, if necessary. In model (III), all 211 particles are mobile and pushed by the growing grains. If a small or large 211 particle is touched by a 123 growing grain, this particle tends to reduce its contact number with the front through a jump on a nearest or also to a next-nearest neighbor liquid site. This rule introduces a dynamically repulsive interaction between the growth front and all 211 particles, whence a pushing/trapping probability (P) transfer matrix-like description for lattice site occupation, - different from model I to III.

These three situations are imagined such that the UCJ process be implemented. As recalled in the introduction, a critical particle size exists below which the particle is pushed by the front and above which the particle is trapped in the crystal matrix. Thus

(I)   corresponds to a critical particle size smaller than the size of the 211 particles;

(II)  corresponds to a critical 211 particle size between the sizes of the small and the large particles considered in the model;

(III) is the case for which the critical size of the 211 particles is high.

The selection, reaction and repulsion process is repeated until the number of possible growth sites is zero. In fact, the above rules impose a stoichiometric balance in the system. Notice that these rules lead to a possible change in the size distribution of the 211 particles during the growth. Finally, it should be pointed out that the above processes can be described along theoretical lines of approach[20], i.e. a "*kinetic growth-with chemical reaction-and pushing trapping mechanism-transfer matrix method*"", involving the three matrices, G, R and P.

4.   RESULTS

Our study was restricted to 123/211 composites with a possible initial 211 excess. The conditions of a too low fraction of 211 particles are not of apparent interest. For each dynamical case (II and III), it has been found that the fraction of 123 phase decreases slowly with the number of initial nuclei *n*. By increasing *n*, the length of grain boundaries increases indeed, whence there are more possibilities for unreacted phases to jam between different grains.

For model I, a fraction of about 86% of 123 phase is produced and many

211 particles are trapped inside the grains. The reaction cannot be fully complete and some liquid segregation is observed sometimes inside the grains. The majority of these liquid segregations are found to be located on the grain boundaries. For model II, more 123 phase is produced (about 90%) and the quantity of resulting 211 particles is lower than for model I. Unreacted liquid phases are often located on the grain boundaries. The inner grain structure is cleaner than in model I. For model III, about 93% of 123 phase is produced with few undissolved particles left over. A small fraction of unreacted L phase is nevertheless observed. Segregations of L phase and large 211 particles are located along the grain boundaries.

In the three models, the remaining 211 particles trapped in the 123 phase are mainly the large ones. Only a few small 211 particles are left over in the grains or at the grain boundaries. In other words, fine 211 particles are consumed easily by the process.

An interesting and crucial conclusion consists in stressing that the fraction of resulting phases not only depends on the concentration of yttrium in the initial melt but also depends strongly on the 211 size distribution. In most simulated cases, the whole liquid does not react since it can be trapped between grains owing to some 211 deficiency in the neighborhood. Another major observation has to be pointed out: there is a substantial effect of the dynamical interaction between the 211 particles and the grains in the production of the 123 phase. The 211 pushing phenomenon has a 'positive effect' in the production of the 123 phase. This production appears less dependent on the 211 size distribution if a repulsive interaction exists between the 211 particles and the front.

The fraction of liquid phase remaining in the sample after the growth has been calculated as a function of the initial 211 concentration. The optimum cases, for which there is a minimum fraction of liquid remaining in the sample, are found in a region in the $s_{211}$; $l_{211}$ diagram, corresponding to an excess of about 20% of 211 phase added to a melt of 123 initially stoichiometric composition.

## 5. CONCLUSIONS

After the description of bulk 123-YBaCuO growth, and the relevant features of the microstructure, the possible optimization of such peritectically grown ceramics, from the 123 density point of view in presence of 211 particles, has been recalled through a list of tuning phenomena. At the mesoscopic scale following an extended Eden growth-like conditional model the microstructure complexity can be described taking into account that 211 particles are rejected by the solidifying matrix during the 123-YBCO growth. The model, thus including realistic constraints, like a peritectic reaction and a crystal growth anisotropy leads to microstructures quite similar to those observed in the laboratory including different dispersions of particles ahead and behind the front.

The simulations put into evidence the effect of the 211 size distribution itself on the resulting microstructure. Even though, the model considers only two different 211 sizes, the model predicts that the refining of the 211 particles leads to better samples. The optimum situation, i.e. minimization of the fraction of remaining liquid phase, is found to occur for a 20% excess of initial 211 phase.

The above results show the great importance of the size of the 211

particles in controlling the microstructure of 123/211 composites. This effect of the refining of 211 particles on the microstructure was effectively observed in various experimental studies.

ACKNOWLEDGMENT

This work has been partially supported through Action de Recherches Concertée Programs of the University of Liège ARC 94-99/174 and ARC 02/07-293. MA and RC thank also RW.0114881-VESUVE program for other partial support.